\documentstyle[12pt,a4wide]{article}

\newcommand{\be}[1]{\begin{equation}\label{#1}}
\newcommand{\ee}{\end{equation}}
\newcommand{\ba}[1]{\begin{eqnarray}\label{#1}}
\newcommand{\ea}{\end{eqnarray}}
\begin{document}
\hfill    {\bf FUB - HEP/93 - 19}

\bigskip

\bigskip

\bigskip


\centerline{\large\bf MULTIDIMENSIONAL INTEGRABLE}
\bigskip
\centerline{\large\bf COSMOLOGICAL MODELS}
\bigskip
\centerline{\large\bf WITH DYNAMICAL AND SPONTANEOUS}
\bigskip
\centerline{\large\bf COMPACTIFICATION}

\bigskip

\bigskip
\centerline{ \bf U. BLEYER
\footnote{This work was supported by WIP grant 016659}
}
\centerline{WIP-Gravitationsprojekt, Universit\"at Potsdam }
\centerline{An der Sternwarte 16 }
\centerline{ D-14482 Potsdam, Germany}
\bigskip
\centerline{and}
\bigskip
\centerline{\bf A. ZHUK
\footnote{Permanent address: Department of Physics,
University of Odessa,
2 Petra Velikogo,
Odessa 270100, Ukraine}
\footnote{This work was supported in part by DAAD and by DFG grant 436
UKR - 17/7/93}
}
\centerline{WIP-Gravitationsprojekt, Universit\"at Potsdam }
\centerline{An der Sternwarte 16 }
\centerline{ D-14482 Potsdam, Germany}
\centerline{and}
\centerline{Fachbereich Physik, Freie Universit\"at Berlin}
\centerline{Arnimallee 14}
\centerline{ D-14195 , Germany}

\bigskip

\bigskip

\noindent
{\large\bf Abstract}

\bigskip

\noindent
Multidimensional cosmological models with $n (n > 1)$ spaces of
constant curvature are discussed classically and
with respect to canonical quantization. These
models are integrable in the case of  Ricci flat internal spaces.
For positive curvature in the external space we find exact solutions
modelling dynamical as well as spontaneous compactification of
internal spaces.


\newpage

\section{INTRODUCTION}

\noindent
The interest in the Kaluza-Klein idea of geometric
unification of interactions actually initiates a lot of work. On one
side the consistent formulation of supersymmetry or superstring
theory is  possible only in a space-time with more than four
dimensions. In field theory higher dimensions act on four dimensional
space-time through the generation of particle masses
[1 - 3],
or the influence on effective constants of nature (like the constant of
gravitation)
[4 - 7]
Higher dimensions can generate observable remnants like cosmic rays of
ultrahigh energy
\cite{8}.
They are also used for a geometric
interpretation of internal quantum numbers (like electric charge)
\cite{2}.

On the other side multidimensional models are successfully considered
in cosmology. It was shown that
such higher dimensional stages may have significant influence on the
evolution of our external space. Obviosly, the internal dimensions are
not observable at the actual state of the universe. Therefore, all
multidimensional cosmological models (MCM) have to describe
the compactification of the internal dimensions up to the
actual time. As a consequence all MCM  can be devided into two
different classes. The first class consists of models where  from the very
beginning the internal dimensions are assumed to be static
and of the scale of Planck length $L_{Pl} \sim 10^{-33}$ cm.
Such MCM are called {\em models with spontaneous compactification}
[1 - 3],
[9 - 19].
The other class consists of models where the internal dimensions
undergo a dynamical evolution like the external space-time.
But during the evolution of the universe the internal spaces
contract for several orders with respect to the external one. These
are the MCM {\em with dynamical compactification}. In both classes we find
purely gravitational models as well as models with different types of
matter (scalar field, electromagnetic field, Yang-Mills field, ideal
fluid, one-loop quantum corrections etc.).

Of special interest are exact solutions
(
[5, 14, 18],
[20 - 30]
)
because they can be used for a detailed study of the evolution of our
space, of the compactification of the internal space and of the
behaviour of matter fields.

One of the most natural  MCM generalizing the
Friedmann-Robertson-Walker (FRW) universe is given by a toy model with
the topology $R \times M_{1} \times \dots \times M_{n}$ where $M_{i}
(i = 1, \dots , n)$ denote spaces of constant curvature. One of
these spaces, say $M_{1}$, describes the exterior space but all the others
are internal spaces. This model was considered especially in papers
[31 - 40].

In
[31]
there was given  a gauge covariant Wheeler-DeWitt equation (WDW)  for
this model. As shown in
[37]
the classical field equations can be integrated for this model, if at
most one of the spaces $M_{i}(i = 1, \dots , n)$ is non Ricci flat
\footnote{The criterion of integrability given in
[37]
seems to be sufficient. To decide its necessity in the case with more than
one non Ricci flat spaces has to be studied in more detail.}.
In the quantum theory the WDW equation can be integrated also in
this case
[34 - 36, 38].
This integrability takes place also for a MCM with only Ricci flat
spaces and a non vanishing cosmological constant
[39]
or filled with ideal fliud
[32,40].

In the present paper we consider the integrable case of a MCM where
the non Ricci flat space is of constant positive curvature. The main
aim consists in the study of the compactification in this model,
spontaneous as well as dynamical one. If the Ricci flat
spaces describe the internal spaces they are assumed to be compact. This can
be ensured by appropriate boundary conditions. The most simple
example of such Ricci flat spaces is given by d-dimensional tori. As a
matter source we consider a homogeneous scalar field minimally coupled
to gravity. We shall analize the case of a free scalar field as well
as  a field with a special potential. In all these cases the models
considered are integrable.

It will be shown that the MCM investigated here has solutions
describing spontaneous and dynamical compactification. The paper is
organized as follows. In sec.2 we describe the MCM and represent it
in appropirate coordinates. Sec. 3 is devoted to dynamical
compactification, sec. 4 to the spontaneous one. Conclusions and an
extended list of references complete the paper.


\section{MULTIDIMENSIONAL  COSMOLOGICAL \newline  MODELS}


Let us start with a $D$-dimensional space-time manifold
\be{1}
M = R \times M_{1} \times \dots \times M_{n}
\end{equation}
containing $n$ spaces of constant curvature $M_{i}$ with $d_{i}$ dimensions.
$M_{1}$ denotes our three dimensional space, but we do not restrict
ourself to this special value and use $d_{1} > 1$ as dimension.
In general the line element is given by
\be{2}
ds^{2} = g_{AB}dx^{A}dx^{B} \hspace{2cm} A, B = 0, \dots, D-1
\end{equation}
We consider the case of homogeneous  spaces where
the line element is given by
\be{3}
ds^{2} = - dt^{2} exp(2\gamma (t)) + \sum_{i=1}^{n}
a_{i}^{2}(t)  g_{(i)}
\end{equation}
and the $a_{i}= e^{\beta^{i}}$ denote the scale factors of the
different factor spaces. The metrics of these spaces are
given by
\be{4}
g_{(i)} = h_{m_{i}n_{i}}dx^{m_{i}}dx^{n_{i}} \hspace{1cm} m_{i},
n_{i} = 1, \dots, d_{i}
\end{equation}
With the demand the $M_{i}$ to be  spaces of constant curvature we get
\be{5}
g_{(i)} = \frac{\sum (dx^{n^{i}})^{2} }{(1 + \frac{1}{4}
k_{i} \sum (x^{n^{i}})^{2})^{2}  }
\end{equation}
where $k_{i} = 0, \pm 1$. The scalar curvature of $M_{i}$ is given by
\be{6}
R[g_{(i)}] \equiv \theta_{i} = k_{i}d_{i}(d_{i} - 1)
\ee

Matter sources compatible with this symmetry can be described  by a
homogeneous matter field or phenomenologically  by an energy momentum
tensor of a generalized ideal fluid. We consider the case of a
minimally coupled homogeneous scalar field $\varphi$ with the potential
$V(\beta, \varphi)$ depending
on $\varphi$ as well as on  the $\beta^{i}$. This gives us the possibility  to
investigate models with an arbitrary scalar field potential
$V(\beta,\varphi)\equiv U(\varphi)$ as well
as (for $\varphi =$ const) models with an arbitrary potential
$V(\beta,\varphi)\equiv V(\beta)$. Effective
potentials of the form $V(\beta)$ may have their origin in an ideal fluid
matter source [32,40], for compact inner spaces in the ''Casimir
effect'' [11] or in the ''monopole''ansatz [41]. At last the general
form of the potential $V(\beta,\varphi)$ leads us to new integrable models.
An example of this kind of potentials will be presented here.

In the case of a free scalar field and only one of the spaces being
non Ricci flat the multidimensional models fulfil a general condition of
integrability
[37].
In what follows we restrict our consideration to the case
$\theta_{1} > 0$ and $\theta_{i} = 0$ for all $i = 2, \dots, n$.
We have, therefore, a special class of integrable models.
The compactness of the flat internal spaces has to be ensured by
apropriate boundary conditions.

The  action $S$ for the model with the metric
(\ref{3}) and a minimally coupled homogeneous scalar field can be written
in the form
[31]
\be{7}
S = \int {\cal L} d t
\end{equation}
where the Lagrangian reads
\ba{8}
{\cal L} & = & \frac{\mu}{2}e^{-\gamma + \sum_{i=1}^{n} d_{i}\beta^{i} }
\left \{\sum_{i=1}^{n} d_{j} (\dot\beta^{j})^{2} - (\sum_{i=1}^{n} d_{j}
\dot\beta^{j})^{2} + \kappa^{2}\dot\varphi^{2}\right \} \nonumber \\
& & + \frac{\mu}{2}  e^{\gamma + \sum_{i=1}^{n} d_{i}\beta^{i}}
\theta_{1}e^{-2\beta^{1}}
- \mu\kappa^{2} V(\beta, \varphi) e^{\gamma + \sum_{i=1}^{n} d_{i}\beta^{i} }
\ea
Here $\kappa^{2}$ denotes the gravitational constant and $\mu =
\prod_{i=1}^{n} V_{i}/\kappa^{2}$ where $V_{i}$ is the volume of
$M_{i}$: $V_{i} = \int_{M_{i}} d^{d_{i}} y
(det(g_{m_{i}n_{i}}))^{1/2}$. The metric (\ref{3}) can be normalized
in such a way that $\mu = 1$. For example, this can be done by a transformation
$\gamma \rightarrow \gamma + \frac{1}{2}\left( 1 - \frac{D}{2}
\right)^{-1}\ln\mu, \beta^{i} \rightarrow \beta^{i} +
\frac{1}{2}\left( 1 - \frac{D}{2}
\right)^{-1}\ln\mu $.  In what follows we use this property.
Further we use natural units with $\kappa^{2} = 1$.

The model will be analized in two different time gauges. In the
harmonic time gauge
[31]
with the time
coordinate $\tau$ we have $\gamma = \sum_{i=1}^{n} d_{i}\beta^{i}$,
and in the synchronous time gauge with coordinate $t$ we have
$\gamma = 0$.

The free choice of the time coordinate leads to a constraint equation
which reads
\be{9}
e^{-\gamma}
\left[ \sum_{i=1}^{n} d_{i} (\dot\beta^{i})^{2} - (\sum_{i=1}^{n} d_{i}
\dot\beta^{i})^{2} + \dot\varphi^{2} \right]
- e^{ \gamma}\theta_{1}e^{-2\beta^{1}}
+ 2 V(\beta, \varphi) e^{\gamma} = 0
\ee

The given cosmological model was already considered earlier in
[34 - 36, 38],
where the main aim was the analysis of the quantized system. In the
present paper we intend to consider mainly problems connected with
compactification in Kaluza-Klein theory with respect to the given model.
The model under consideration belongs to the class of integrable
models what makes a detailed consideration of these problems possible.

Let us first consider the models where no scalar potential is
present: $V(\beta, \varphi) = 0$.
It was shown in
[34 - 36, 38],
that the field equations for this model can be integrated most easily
using the following coordinates:
\ba{10}
q v^{0} & = & (d_{1} - 1) \beta^{1} + \sum_{i=2}^{n} d_{i}\beta^{i},
                   \nonumber \\
q v^{1} & = & \left[(D -2)/(d_{1} {\sum}_{2} \right]^{\frac{1}{2}}
             \sum_{i=2}^{n} d_{i}\beta^{i},  \\
q v^{i} & = & \left[(d_{1} -1)d_{i}/(d_{1} {\sum}_{i}{\sum}_{i+1}
            \right]^{\frac{1}{2}}
             \sum_{j=i+1}^{n} d_{j}(\beta^{j} - \beta^{i}),
          \hspace{1cm} i = 2, \dots, n-1 \nonumber
\end{eqnarray}
Here we used the notations $D = 1 + \sum_{i=1}^{n} d_{i}$, $q^{2} =
(d_{1} - 1)/d_{1}$, and ${\sum}_{i} = \sum_{j=i}^{n} d_{j}$.
The inverse transformation is given by

\ba{11}
(d_{1} - 1) \beta^{1} & = & q v^{0}  - \left[ \frac{ \sum_{2} (d_{1} -
1) }{D - 2} \right]^{\frac{1}{2}}v^{1},
                   \nonumber \nopagebreak \\
\beta^{i+1} & = & \left[\frac{d_{1} - 1}{(D -2){\sum}_{2}}
            \right]^{\frac{1}{2}}v^{1}
            + \sum_{k=2}^{i} \left[
\frac{d_{k}}{{\sum}_{k}{\sum}_{k+1}} \right]^{\frac{1}{2}}v^{k}
\nonumber  \\
&  &  - \left[
\frac{\sum_{i+2}}{d_{i+1}{\sum}_{i+1}} \right]^{\frac{1}{2}} v^{i+1},
\qquad i = 1, \dots, n-2 \\
\beta^{n} & = & \left[\frac{d_{1} - 1}{(D -2){\sum}_{2}}
            \right]^{\frac{1}{2}}v^{1}
          +  \sum_{k=2}^{n-1} \left[
\frac{d_{k}}{{\sum}_{k}{\sum}_{k+1}} \right]^{\frac{1}{2}}v^{k}
              \nonumber
\end{eqnarray}
The field equations take the form of equations of motion for the
dynamical variables $v^{i}$ of a dynamical system with $n$ degrees of
freedom. In the harmonic time gauge the solutions read
[35, 36, 38]
\be{12}
e^{qv^{0}} = \frac{\sqrt{\epsilon/ \theta_{1}}}{\cosh \left[
q\sqrt{\epsilon} \tau \right]},
\hspace{1cm} - \infty < \tau < + \infty
\ee
and
\be{13}
v^{i} = \nu^{i} \tau + c^{i}, \hspace{1cm} i = 1, \dots, n-1
\ee
for the $v^{i}$ and for the scalar field
\be{14}
\varphi =  \nu^{n} \tau + c^{n}
\ee
Here we used
\be{15}
\epsilon = \sum_{i=1}^{n} (\nu^{i})^{2} > 0
\ee
and the
$\nu^{i}, c^{i}, i= 1, \dots, n$ are constants of integration.
In minisuperspace of vectors $\vec v = (v^0, v^1, \dots,
v^{n-1},v^n\equiv \varphi)$ the indices are raised and lowered by the
diagonal metric $\eta = (-1, +1, \dots,+1)$ \cite{34}. Thus, we have
$v^0=-v_0, v^i=v_i, \nu^i=\nu_i$ and $c^i=c_i, i=1, \dots, n$.

We can generalize our model to the case of Einstein spaces $M_i$ with
$R[g_{i}] = \lambda_id_i$ instead of (\ref{6}). The $\lambda_i$ are
arbitrary constants. In order to get the forthcoming solutions for
this general case we have to perform the substitution $\theta_1
\rightarrow \lambda_1d_1$ in (\ref{12}). In chapters 3 and 4 we
sometimes use the relation $\theta_1=d_1(d_1-1)$ valid for a space
$M_1$ of positive constant curvature ($k=+1$). But all formulas
containing this relation can be trivially rewritten for an
Einstein space $M_1$.

{}From (\ref{11}), (\ref{12}), and (\ref{13}) it is easy to get the
explicit expressions for the scale factors
$a_{i} = e^{\beta^{i}}$ in the harmonic time gauge for arbitrary $n
\ge 2$. These formulas show that the dynamical behaviour of the
universe is very complicated in the case of arbitrary $n$. Some of the
factor spaces may expand and others contract at the same time. It
depends on the signs of the constants $\nu_{i}$ and relationships
between coefficients in formula  (\ref{11}). The general analysis with
arbitrary $\nu_{i}$ and $d_{i}$ is hardly possible to perform. Each
choice should be consedered as a separate case. We shall show how to
do this on the example of two particular cases. First of all we shall
consider the very popular among cosmologists two-component model, $n =
2$, i.e. the model where we have only one
internal space in addition to the external one and, therefore, only
two scale factors. In this case it is possible to perform the general
analysis for arbitrary $\nu_{1}$, $\nu_{2}$ and $d_{1}$, $d_{2}$.
Here, we shall give the explicit expressions for the scale factors in
the synchronous time widely used in cosmology.
We shall show for this case the occurence of dynamical
compactification. This means a cosmological development where one
scale factor monotonically increases while the other one remains on
a much smaller scale.

Another important particular case which will be considered here is the
case with spontaneous compactification of the inner factor spaces.
This case will be analized for arbitrary $n \ge 2$ and the explicit
expresiion for the scale factor of the external (our) space in the
synchronous system will be obtained also.

\section{DYNAMICAL COMPACTIFICATION}

For the case of a two-component cosmological model, i.e. with $n = 2$
in (\ref{11}) - (\ref{15}) we can easyly get an
expression for the scale factors in harmonic time gauge. We find
$a_{1,2}$ as functions of the harmonic time $\tau$
\be{16}
a_{1}^{d_{1} - 1} = \frac{2 {a_{(0)}}_{1}^{d_{1} - 1}}
{e^{\sqrt{\frac{d_{2}(d_{1} - 1)}{D - 2}}\nu_{1}\tau}
\left[ e^{\sqrt{\frac{(d_{1} - 1)(\nu_{1}^{2} + \nu_{2}^{2})}{d_{1}}}\tau}
  +    e^{- \sqrt{\frac{(d_{1} - 1)(\nu_{1}^{2} + \nu_{2}^{2})}{d_{1}}}\tau}
\right]}
\ee
\be{17}
a_{2}^{d_{2}} = {a_{(0)}}_{2}^{d_{2}} e^{\sqrt{\frac{d_{2}(d_{1} -
1)}{D - 2}}\nu_{1}\tau}
\ee
Here ${a_{(0)}}_{1} $ and ${a_{(0)}}_{2} $ are connected with the
constant of integration $c_{1}$ by the expressions
\be{18}
{a_{(0)}}_{1}^{d_{1} - 1} = \sqrt{\frac{\nu_{1}^{2} +
\nu_{2}^{2}}{d_{1}(d_{1} - 1)}} e^{- \sqrt{\frac{d_{2}(d_{1} -
1)}{D - 2}}c_{1}}
\ee
\be{19}
{a_{(0)}}_{2}^{d_{2}} =  e^{\sqrt{\frac{d_{2}(d_{1} -
1)}{D - 2}}c_{1}}
\ee
{}From this we get the connection between ${a_{(0)}}_{1}$ and
${a_{(0)}}_{2}$
\be{20}
{a_{(0)}}_{1}^{d_{1} - 1}{a_{(0)}}_{2}^{d_{2}} = \sqrt{\frac{\nu_{1}^{2} +
\nu_{2}^{2}}{d_{1}(d_{1} - 1)}}
\ee
We have three different types of development of the scale factors $a_{1}$
and $a_{2}$ in dependence from the relation between the constants
$\nu_{1,2}$ and the dimensions of the spaces. Let us consider these
cases in more detail.

\bigskip
\noindent
1. $\hspace{1cm}  \sqrt{d_{1}d_{2}\nu_{1}^{2}} - \sqrt{(d_{1} + d_{2}
- 1)(\nu_{1}^{2} + \nu_{2}^{2})} > 0$
\bigskip

In this case the scale factors  $a_{1}$ and $a_{2}$ are permanently in
opposite phase: Either  $a_{2}$ is contracting from $\infty$ to $0$
and $a_{1}$ is expanding from $0$ to $\infty$ (for $\nu_{1} < 0$), or
$a_{2}$ increases from $0$ to $\infty$
and $a_{1}$ decreases  from $\infty$ to $0$ (for $\nu_{1} > 0$). This
type of behaviour can take place also in the case where no scalar
field is present ($\nu_{2} = 0)$. It is clear that condition 1. is not
valid for $d_{2} = 1$, if only $\nu_{2}$ is not imaginary.

\bigskip
\noindent
2. $\hspace{1cm}  \sqrt{d_{1}d_{2}\nu_{1}^{2}} - \sqrt{(d_{1} + d_{2}
- 1)(\nu_{1}^{2} + \nu_{2}^{2})} < 0$
\bigskip

Also in this case we have two types of possible behaviour depending on the
sign of $\nu_{1}$. While the scale factor $a_{2}$ is contracting from
$\infty$ to $0$ the scale factor $a_{1}$ expands up to a maximal value
and then starts to shrink up to $0$ (for $\nu_{1} < 0)$. On the other
hand  $a_{2}$ increases from $0$ to $\infty$ while
$a_{1}$ increases  up to a maximal value and contracts  to $0$ (for
$\nu_{1} > 0$). Solutions with analogical  behaviour were described
earlier in
[24 - 26],
Obviosly, this case can not take place for vanishing scalar field
($\nu_{2} = 0$). In the classical Kaluza-Klein case with $d_{2} = 1$
the condition 2. is satisfied for any real $\nu_{2}$.

\bigskip
\noindent
3. $\hspace{1cm}  \sqrt{d_{1}d_{2}\nu_{1}^{2}} - \sqrt{(d_{1} + d_{2}
- 1)(\nu_{1}^{2} + \nu_{2}^{2})} = 0$
\bigskip

In this case we find a connection between the constants of
integration $\nu_{1}$ and $\nu_{2}$:
\be{21}
\nu_{2}^{2} = \left( \frac{d_{1}d_{2}}{d_{1} + d_{2} -1} - 1
\right)\nu_{1}^{2}
\ee
{}From this expression we can see that $d_{2} = 1$ (the classical
Kaluza-Klein assumption) corresponds to the case with vanishing  scalar
field. Using (\ref{21}) the expression (\ref{20}) takes the form
\be{22}
{a_{(0)}}_{1}^{d_{1} - 1}{a_{(0)}}_{2}^{d_{2}} = \sqrt{\frac{d_{2}}
{(d_{1} - 1)(d_{1} + d_{2} - 1)}} \mid \nu_{1} \mid
\ee
The dependence of the scale factors on harmonic
time reads
\be{23}
a_{2}^{d_{2}} = {a_{(0)}}_{2}^{d_{2}} e^{\sqrt{\frac{d_{1} - 1}{d_{2}(D -
2)}}\nu_{1}\tau }
\ee
\be{24}
a_{1}^{d_{1} - 1} = {a_{(0)}}_{1}^{d_{1} - 1}\left( 1 \pm \tanh
{\sqrt{\frac{d_{2}(d_{1} - 1)}{(D -
2)}} \mid \nu_{1} \mid \tau} \right)
\ee
where the upper sign corresponds to the case $\nu_{1} < 0$ and the
lower sign to  $\nu_{1} > 0$.
The behaviour of the scale factors is shown in fig.~1 and fig.~2. We can see
that the phenomenon of dynamical compactification takes place for
$\tau > 0$ if $\nu_{1} > 0$ and for $\tau < 0 $ if $\nu_{1} < 0$.

The simple form of eqns. (\ref{23}), (\ref{24}) makes it
possible to find the explicit connection between harmonic and
synchronous time coordinates. We have the general differential coonection
\be{25}
dt = \pm e^{\gamma}d\tau = \pm a_{1}^{d_{1}}a_{2}^{d_{2}} d\tau
\ee
We get finally
\be{26}
t = \pm \tilde{c} \int \frac{dy}{(y^{2} + 1)^{\frac{d_{1}}{d_{1} -
1}}}  + \tilde{\tilde c}
\ee
where
\be{27}
y = e^{\pm \sqrt{\frac{d_{2}(d_{1} - 1)}{D - 2}} \mid \nu_{1} \mid \tau}
\ee
(in eqn. (\ref{27})  the positive sign corresponds to $\nu_{1} > 0$,
the negative
one to $\nu_{1} < 0$) and
\be{28}
\tilde{c} = 2^{\frac{d_{1}}{d_{1} - 1}} \,  \sqrt{\frac{D - 2}{d_{2}(d_{1} -
1)}} \frac{1}{\mid \nu_{1} \mid}
{a_{(0)}}_{1}^{d_{1}}{a_{(0)}}_{2}^{d_{2}}
\ee
Changing the origin of the synchronous time we can achieve
$\tilde{\tilde c} = 0$

Let us consider some special cases. If we have $d_{1} = 2, d_{2} \geq
1$ then (\ref{26}) integrates to give
\be{29}
t = \pm \frac{\tilde c}{2} \left( \frac{y}{y^{2} + 1} + \arctan y
\right), \hspace{1cm} \mid t \mid \le \frac{\tilde c \pi}{4}
\ee
This case is interesting in the sense that for $d_{2} = 1$ this model
describes a three dimensional anisotropic Kantowski-Sachs universe
\cite{42}
without scalar field.

A simple  expression for the scale factors we find in the most
realistic Kaluza-Klein case $d_{1} = 3, d_{2} \geq 1$:
\be{30}
t = \pm \tilde c \frac{y}{(y^{2} + 1)^{\frac{1}{2}}}, \hspace{1cm}
\mid t \mid \le \tilde c
\ee
and with this time coordinate the solution reads
\ba{31}
a_{1} & = & 2^{\frac{1}{2}}{a_{(0)}}_{1}\left[ 1 -
{\left(\frac{t}{\tilde c} \right)}^{2} \right]^{\frac{1}{2}}  \nonumber \\
a_{2} & = & {a_{(0)}}_{2}{\left[ \frac{t^{2}}{{\tilde c}^{2} - t^{2}}
\right]}^{\frac{1}{2d_{2}}}          \\
\varphi  & = & \pm \frac{1}{2}\sqrt{\frac{d_{2} - 1}{d_{2}}}\ln \left(
\frac{t^{2}}{{\tilde c}^{2} - t^{2}} \right)  \nonumber
\end{eqnarray}
{}From this solution we see that the point $t = 0$ is a point of
maximum for the scale factor $a_{1}$ (turning point) and a point of
minimum for $a_{2}$ (point of repulsion). The derivative $da_{2}/dt$
is not smooth at $t = 0$.
The scale factors $a_{1}$ and $a_{2}$ are permanently in opposite
phase and always exist time intervals where dynamical compactification
is realized. On the one hand $a_{1}$ starts to increase from zero at
$t = - \tilde c$
to some maximum at $t = 0$.  At the same time
$a_{2}$ shrinks from $\infty$ ($t = - \tilde c$) to zero ($t = 0$).
This behaviour corresponds to $\nu_{1} < 0$. On the other hand, for
$\nu_{1} > 0$ $a_{1}$ monotonically decreases from its maximal value
at $t = 0$ to zero at $t = + \tilde c$ and $a_{2}$ increases from
zero ($t = 0$) to $\infty$ ($t = + \tilde c$)(fig.~3).
The total space volume
is proportional to $V_{3 + d_{2}} = a_{1}^{3}a_{2}^{d_{2}} \sim
\mid t \mid ({\tilde c}^{2} - t^{2}) \rightarrow 0$ if
$t \rightarrow 0, \pm \tilde c$.

It is easy to see that eq. (\ref{31}) fulfils the constraint equation
(\ref{9}) for $\tilde c = \sqrt 2 {a_{(0)}}_{1}$. Using the
definition for $\tilde c$ (\ref{28}) this relation can be also written in the
form ${a_{(0)}^{2}}_{1}{a_{(0)}^{d_{2}}}_{2} = \left( d_{2}/2(d_{2} +
2)  \right)^{1/2} \mid\nu_{1}\mid $ what coincides with expression
(\ref{22}) for $d_{1} = 3$.

{}From eq. (\ref{31}) it can be seen that the scalar field changes
monotonically from $- \infty$ to $+ \infty$ as a result of the
vanishing potential $(V(\beta, \varphi) = 0)$. But taking into account a
nonvanishing potential of the scalar field in the Lagrangian (\ref{8})
it is in general not possible to separate variables and to integrate
the equations.

There is one comparatively easy case where this problem can be
overcome. Let us consider the potential
\be{32}
V(\varphi, \beta) = U(\varphi)e^{- 2 \sum_{i=1}^{n}d_{i}\beta^{i}}
\ee
Than we find for the Lagrangian (\ref{8}) in $v$-coordinates and in
the harmonic time gauge the expression
\be{33}
{\cal L} = \frac{1}{2}(\eta_{ik}\dot v^{i}\dot v^{k} + \dot \varphi^{2}) +
\frac{1}{2} \theta_{1}e^{2qv_{0}} - U(\varphi)
\ee
where $\eta_{ik} = diag(-, +, + \dots, +)$, The overdot denotes the
derivative with respect to the harmonic time and we put $\theta_{i} =
0, i = 2, \dots, n$.
In this case we have the possibility to separate the variables and the
equations of motion take the form
\pagebreak
\ba{34}
\ddot v^{0} + \theta_{1}qe^{2qv^{0}} & = & 0 \nonumber \nopagebreak \\
\ddot v^{i} & = & 0, \qquad i= 1, \dots, n-1 \nopagebreak \\
\ddot\varphi + \frac{\partial U}{\partial \varphi} & = & 0 \nonumber
\end{eqnarray}
The constraint equation (\ref{9}) can be written as
\be{35}
(\dot v^{0})^{2} + \theta_{1} e^{2qv^{0}}  = \epsilon
\ee
where
\ba{36}
\epsilon & = & \sum_{i = 1}^{n - 1}(\dot v^{i})^{2} + \dot\varphi^{2} + 2
U(\varphi) \nonumber \\
& = &  \sum_{i = 1}^{n - 1}(\nu_{i})^{2} + \dot\varphi^{2} + 2 U(\varphi)
\ea
As before $\nu_{i}$ denotes the constants of integration of the
equations $\ddot v^{i} = 0 (i = 1, \dots, n - 1)$.

Let us now restrict our consideration to a two-component universe (one
internal space) and let us take the potential $U(\varphi) =
\frac{1}{2}m^{2}\varphi^{2}$. Then the potential $V(\beta, \varphi)$ can be
rewritten in the form $V(\beta, \varphi) = \frac{1}{2}m^{2}\Phi^{2}$,
where $\Phi = \varphi/V_{D'}$ and $V_{D'} = a_{1}^{d_{1}}a_{2}^{d_{2}}$
is proportional to the voume of the $D' = d_{1} + d_{2}$  dimensional space.
For $\varphi = const$ we have $V(\beta, \varphi)  \sim
1/(a_{1}^{d_{1}}a_{2}^{d_{2}})^{2}$. This expression is similar to the
monopol ansatz of Freund and Rubin
\cite{41}
but with the difference that their monopole is assumed in the internal
space only, but our scalar field is given in the whole universe.
Therefore, we have in the expression for $V(\beta, \varphi)$ an
additional factor $a_{1}^{2d_{1}}$.

With the choice $U(\varphi) = \frac{1}{2}m^{2}\varphi^{2}$ we find for eq.
(\ref{34}) the solution for the scalar field
\be{37}
\varphi = \varphi_{0} \cos [m (\tau - \tau_{0})]
\ee
where $ \varphi_{0}$ and $ \tau_{0}$ are constants of integration.

It can be seen from (\ref{34} - \ref{37}) that the behaviour of the
scale factors will be the same as in the case $U(\varphi) = 0$ if we
put $\nu_{2}^{2} = m^{2}\varphi_{0}^{2}$, because we have once more
$\epsilon = \nu_{1}^{2} + \nu_{2}^{2} = const$. Therefore, for this
choice of the potential $U(\varphi)$ we have the same picture for
compactification as before. The most simple expression for the scalar
field in synchronous time can be obtained for the case 3. (i.e.
$\sqrt{d_{1}d_{2}\nu_{1}^{2}} = \sqrt{(d_{1} + d_{2} - 1)(\nu_{1}^{2}
+ \nu_{2}^{2})}$) if $d_{1} = 3$. Then, we get \be{38} \varphi =
\varphi_{0} cos\left[ \frac{m}{2 \nu_{1}} \sqrt{\frac{d_{2} +
2}{2d_{2}}}\ln \frac{t^{2}}{A(\tilde c^{2} - t^{2})} \right] \ee where
$\tilde c$ is defined by eqn. (\ref{28}), $A = \ln
\frac{t_{0}^{2}}{\tilde c^{2} - t_{0}^{2}}$ and the constant $t_{0}$
corresponds to the harmonic time $\tau_{0}$ (using eqn. (\ref{30}). We
can see that the scalar field oscillates. At $t = 0, \pm \tilde c $
these oscillations reach infinite frequency.

At the quantum level we have to replace the quadratic
amplitude $\varphi_{0}^{2}$ in the expression for $\epsilon$ by
discrete energy levels. The quantization procedure changes the
constraint equation (\ref{35}) into the Wheeler-DeWitt equation
\be{39}
\left(
- \frac{\partial^{2}}{\partial {v^{0}}^{2}} + \theta_{1} e^{2q v^{0}}
+ \frac{\partial^{2}}{\partial {v^{1}}^{2}} +
\frac{\partial^{2}}{\partial \varphi^{2}} - m^{2} \varphi^{2}
\right) \Psi = 0
\ee
which has the solution
\be{40}
\Psi_{\nu_{1},n} = e^{i\nu_{1} v^{1}} e^{- \frac{m}{2} \varphi^{2}}
H_{n}(\varphi \sqrt{m}){\rm C}_{i\sqrt{\epsilon}/q}\left(
\frac{\sqrt{\theta_{1}}}{q} e^{qv^{0}} \right)
\ee
where $H_{n}$ are Hermite polynomials and ${\rm C} = {\rm I}, {\rm K}$
denotes the modified Bessel functions. Here we have
\be{41}
\epsilon = \nu_{1}^{2} + (2n + 1)m, \qquad n = 0, 1, 2, \dots; \qquad
- \infty < \nu_{1} < + \infty
\ee

In conclusion of this chapter we would like to mention the existence
of a special case of n-component cosmological model $(n > 2)$, which
can be formally reduced to the two-component case described above.
This special case will be given by the following special choice of the
constants of integration in eqn. (\ref{13})
$$
\nu_{1} \neq 0, \hspace{1cm} \nu_{2} = \dots = \nu_{n-1} = 0  \nonumber
$$
Then we find from the coordinate transformation (\ref{11})
\be{42}
a_{i} = e^{B_{i}}a_{2},  \hspace{1cm}  i = 3, \dots , n
\ee
where the $B_{i}$ are arbitrary constants and the n-component case is
reduced to the two-component one $(a_{1}, a_{2})$ considered above  by
the replacement of $d_{2}$ in eqns. (\ref{16}) - (\ref{31}) by $\sum_{2} =
\sum_{i = 2}^{n} d_{i}$. In this particular case all factor spaces
$M_{i} = 3, \dots, n$
have the same dynamical behaviour as $M_{2}$ the evolution of which was
described above.


\section{SPONTANEOUS COMPACTIFICATION}


Spontaneous compactification we call  special solutions to the
equations of motion where only one scale factor undergoes dynamical
development and will be connected with the scale factor of the
external (our) space while all internal space scale factors are fixed.
It is usually assumed that these fixed scale factors are of the order
of the Planck length $L_{Pl} \sim 10^{-33}$ cm. These scale factors
are assotiated to internal dimensions not accessible to direct observations.

For the cosmological model under consideration there are solutions of
this type corresponding to the special case $\nu_{i} = 0, (i = 1,
\dots , n-1)$. Then we have from (\ref{11}), (\ref{13}) $\beta^{i} =
const, (i = 2, \dots , n)$, i.e. $a_{i} = e^{\beta^{i}} = a_{(0)i} =
const, (i = 2, \dots , n)$. In this case we have $\epsilon =
\nu_{n}^{2} > 0$ and Lorentzian solutions occur only in the presence
of a real scalar field $\varphi = \nu_{n}\tau + c_{n}$. In the
harmonic time gauge the dynamical scale factor behaves like
[35, 36, 38]
\be{43}
a_{1}(\tau) = \left[ \frac{\sqrt{\epsilon/\theta_{1}}}{c}
\right]^{\frac{1}{d_{1} - 1}} \left\{ \cosh \left[ (d_{1} - 1)
\sqrt{\epsilon/\theta_{1}} \tau  \right]  \right\}^{- \frac{1}{d_{1} -
1}}
\ee
for $- \infty < \tau < + \infty$ and $c =
\prod_{i=1}^{n}a_{(0)i}^{d_{i}}$.

It is convenient to write the metric (\ref{3}) in the gauge of
conformal time $\eta$, which is connected with the harmonic time
$\tau$ by the relation
\be{44}
\cosh \left[ (d_{1} - 1)\sqrt{\epsilon/ \theta_{1}}\tau \right] =
[\sin (d_{1} - 1)\eta]^{-1}, \qquad 0 \le (d_{1} - 1)\eta \le \pi
\ee
Then we have
\be{45}
ds^{2} = a_{1}^{2}(\eta)(-d\eta^{2} + g_{(1)}) + a_{(0)2}^{2}g_{(2)} +
\dots +  a_{(0)n}^{2}g_{(n)}
\ee
and the scale factor $a_{1}$ depends on the conformal time $\eta$ in
the following way
\be{46}
a_{1}(\eta) = \left( \frac{\sqrt{\epsilon/\theta_{1}}}{c}
\right)^{1/(d_{1} - 1)} \left(\sin[(d_{1} - 1)\eta]  \right)^{1/(d_{1} - 1)}
\ee
{}From (\ref{46}) it can be seen that for $d_{1} = 2$ the scale factor
$a_{1}$ behaves like for the closed radiation dominated Friedmann
universe, and for $d_{1} = 3$ like closed Friedmann universe filled
with ultra stiff matter. This example shows once more that extra
dimensions may play the role of a matter source for the external
space-time.

We also give the explicit expression for the metric in the
synchronous reference system:
\be{47}
ds^{2} = - dt^{2} + a_{1}^{2}(t)g_{(1)} + a_{(0)2}^{2}g_{(2)} +
\dots +  a_{(0)n}^{2}g_{(n)}
\ee
The dependence of the scale factor $a_{1}$ on the sychroneous time is
given by
\be{48}
t = \int \frac{a_{1}^{d_{1} -1}da_{1}}{\sqrt{\frac{\epsilon}
{\theta_{1}c^{2}} - a_{1}^{2(d_{1} - 1)}}} + const
\ee
All these formulas (\ref{43}), (\ref{46}), and (\ref{48}) show that
$a_{1}$ expands from zero to it's maximal value $\left(
\frac{\sqrt{\epsilon/\theta_{1}}}{c} \right)^{1/(d_{1} - 1)}$
and shrinks to zero again.
For $ a_{1} \ll \left( \frac{\sqrt{\epsilon/\theta_{1}}}{c}
\right)^{1/(d_{1} - 1)} $ the scale factor asymptotically behaves like
$a_{1} \sim t^{1/d_{1}}$ what corresponds to a closed Friedmann
universe filled with radiation for $d_{1} = 2$ and ultra stiff matter for
$d_{1} = 3$.

In the case $d_{1} = 2$ the integral (\ref{48}) can be expressed by
elementary functions
\be{49}
a_{1} = \left[ t (2\sqrt{\epsilon/ \theta_{1}}/c - t) \right]^{1/2},
\hspace{1cm} 0 \leq t \leq 2\sqrt{\epsilon/ \theta_{1}}/c
\ee
for $d_{1} \ge 3$ it can be expressed by ellyptic integrals. For
instance, in the case $d_{1} = 3$ we have
\be{50}
t = \left[\frac{1}{c}\sqrt{\frac{\epsilon}{\theta_{1}}} \right]^{1/2}
\left\{ \sqrt{2} \left[ E \left( \Psi, \frac{\sqrt{2}}{2} \right)
- {\rm E}
\left(\frac{\sqrt{2}}{2}\right) \right] - \frac{1}{\sqrt{2}}
\left[ F
\left( \Psi, \frac{\sqrt{2}}{2}\right) - {\rm K} \left(\frac{\sqrt{2}}{2}
\right) \right] \right\}
\ee
where
\be{51}
\Psi = \arccos \left[ a_{1}/\left(\sqrt{\epsilon/ \theta_{1}}/c  \right)^{1/2}
 \right]
\ee
and the constant of integration is chosen in such a way that
$a_{1}(t=0) = 0$.
It is clear that a potential of the form (\ref{32}) (with $U(\varphi) =
\frac{1}{2}m^{2}\varphi^{2}$) would
not change the formulas (\ref{43}) - (\ref{51}) where we would have to
put $ \nu_{n}^{2} = m^{2} \varphi_{0}^{2} = const $.

The solution (\ref{43}) has an interesting analytical continuation
into the Euclidean region. This solution has the topology of a
wormhole
[35, 36, 38].
The analytical continuation of the solution (\ref{46}) symmetrical
with respect to $\eta = 0$ (this corresponds to the symmetry with
respect to the throat of the wormhole in the Euclidean region
[43]
) leads to
\be{52}
a_{1}(\eta) = \left(\frac{\sqrt{\epsilon/ \theta_{1}}}{c}
\right)^{\frac{1}{d_{1} - 1}} \left[ \cosh(d_{1} - 1)\eta
\right]^{\frac{1}{d_{1} -1}}, \hspace{1cm} - \infty < \eta < + \infty
\ee

Now we show that this solutions really describes a wormhole, that
means two asymptotically flat regions connected by a throat
\cite{44,45}. To this end we change to a new  ''time'' coordinate $T$ by the
following transformation
\be{53}
a_{1}(\eta) d\eta = \frac{1}{(d_{1} - 1)a_{1}^{d_{1} - 2}} dT
\ee
(we see from (\ref{53}) that the ''time'' $T$ is the synchronous one
in the case $d_{1} = 2$).
We put (\ref{52}) into (\ref{53}) and find
\be{54}
T = \frac{\sqrt{\epsilon/ \theta_{1}}}{c} \sinh (d_{1} - 1)\eta,
\hspace{1cm} - \infty < T < + \infty
\ee
and
\be{55}
a_{1}(T) = \left[ \left(\frac{\sqrt{\epsilon/ \theta_{1}}}{c}\right)^{2} +
T^{2} \right]^{\frac{1}{2(d_{1} - 1)}}
\ee
{}From (\ref{55}) we can see that the throat has the size
$
\left(\frac{\sqrt{\epsilon/ \theta_{1}}}{c}
\right)^{\frac{1}{(d_{1} - 1)}}
$.
In the T-time gauge the metric takes the form
\ba{56}
ds^{2} & = & \frac{dT^{2}}{(d_{1} - 1)^{2}\left[
\left(\frac{\sqrt{\epsilon/ \theta_{1}}}{c} \right)^{2} +
T^{2} \right]^{\frac{d_{1} - 2}{d_{1} - 1}}} +
\left[\left(\frac{\sqrt{\epsilon/ \theta_{1}}}{c} \right)^{2} +
T^{2} \right]^{\frac{1}{d_{1} - 1}} g_{(1)} \nonumber \\
& & + \sum_{i=2}^{n} a_{(0)i}^{2} g_{(i)}
\ea
Asymptotically we find for $T^{2} \gg \left(\frac{\sqrt{\epsilon/
\theta_{1}}}{c} \right)^{2}$
\be{57}
ds^{2} \approx \frac{dT^{2}}{(d_{1} - 1)^{2}
T^{2\frac{d_{1} - 2}{d_{1} - 1}}} +
T^{\frac{2}{d_{1} - 1}} g_{(1)} + \sum_{i=2}^{n}
a_{(0)i}^{2} g_{(i)}
\ee
If we go over to a new coordinate $R = \mid T \mid^{\frac{1}{d_{1} -
1}}$ the asymptotics (\ref{57}) takes the form
\be{58}
ds^{2} \approx  dR^{2} +
R^{2}g_{(1)} + \sum_{i=2}^{n} a_{(0)i}^{2} g_{(i)}
\ee

In this way we get the following result. If $g_{(1)}$ represents the
$d_{1}$ dimensional sphere the Euclidean region has two asymptotic
regions $(T \rightarrow \pm \infty)$ with the topology $R^{d_{1} + 1}
\times M_{2} \times  \dots \times M_{n}$ which are connected by a
throat of the size
$
\left(\frac{\sqrt{\epsilon/ \theta_{1}}}{c}
\right)^{\frac{1}{(d_{1} - 1)}}
$.
This is a wormhole by the definition of this object.

It is useful to express the metric (\ref{56}) in the synchronous
reference system
\be{59}
ds^{2} =  dt^{2} +
a_{1}^{2}(t)g_{(1)} + \sum_{i=2}^{n} a_{(0)i}^{2} g_{(i)}
\ee
where the scale factor $a_{1}$ depends on the synchronous ''time''
coordinate due to the expression
\be{60}
t = \int \frac{a_{1}^{d_{1} - 1}da_{1}}{\sqrt{a_{1}^{2(d_{1} - 1)} -
\frac{\epsilon}{\theta_{1}c^{2}}}} + const
\ee
In the special case $d_{1} = 2$ this integral can be expressed by
elementary functions
\be{61}
a_{1}^{2}(t) = t^{2} + \frac{\epsilon}{\theta_{1}c^{2}}, \hspace{1cm}
- \infty < t < + \infty
\ee
This follows directly from (\ref{56}). We can get (\ref{61}) also by
analytical continuation of the expression (\ref{49}) symmetrical with
respect to $ t= 0 $.
Such type of wormhole solutions was presented by Hawking in
\cite{46}.
For $d_{1} > 2$ the integral (\ref{60}) can by expressed by elyptical
functions. In the case $d_{1} = 3$ we find
\be{62}
t = \left[\frac{1}{c}\sqrt{\frac{\epsilon}{\theta_{1}}} \right]^{1/2}
\left\{ - \sqrt{2} E \left( \Psi, \frac{\sqrt{2}}{2} \right) +
\frac{1}{\sqrt{2}} F \left( \Psi, \frac{\sqrt{2}}{2}\right) \right\}
+ \frac{1}{a_{1}}\left[ a_{1}^{4} - \frac{\epsilon}{\theta_{1}c^{2}}
\right]^{1/2}
\ee
where
\be{63}
\Psi = \arccos \left[\left( \frac{\sqrt{\epsilon/ \theta_{1}}}{c}
\right)^{1/2} /a_{1} \right]
\ee
Such wormholes were studied by Giddings and Strominger in
\cite{47}.


\section{CONCLUSIONS}

We considered integrable multidimensional cosmological models (MCM)
with $n (n> 1)$ spaces of constant
curvature. Only one space was of positive curvature, all the others
were assumed to be Ricci flat. As a matter source we introduced a
minimally coupled homogeneous scalar field as a free field or with
a special type of potential. It was shown that this models posses
solutions describing the process of dynamical compactification of internal
dimensions as well as spontaneous compactification. Moreover,
dynamical compactification takes place in the presence of a scalar field
and for pure gravity. For spontaneous compactification the
presence of a real scalar field in the Lorentzian region is a
necessary condition.
For the case with an imaginary
scalar field or without any scalar field our solution belongs
completely to the Euclidean region. In the case of a real scalar field
the solutions with spontaneous compactification permit an interesting
continuation to the Euclidean region describing Euclidean wormholes.

For the quantized model  the Wheeler-DeWitt (WDW) equation
was analyzed for the case
with a scalar field potential like described above. The corresponding
exact solutions were found. The WDW equation with a free scalar field
was considered earlier in
[34 - 36, 38]
where the exact solutions were presented and analyzed. In this case a
special class of solutions satisfies the boundary conditions  of quantum
wormholes
\cite{48}.

Of interest is the case where the non Ricci flat space is of negative
curvature. It is easy to see from the field equations, that in this
case we have a more rich situation than in the case described in this
paper. There are solutions which belong completely to the Lorentzian
region which do not possess analytical continutation to the Euclidean
one and the
solutions which have such a continuation. Usually, the last class is
connected with models of tunneling universes. The consideration of
this case and the analysis of the compactification problem in it will
be considered in a separate paper.

\bigskip

\noindent
{\bf Acknowledgment}
The work was sponsored by the WIP project
016659/p. One of us (A. Z.) was supported by DAAD and by DFG grant
436UKR-17/7/93.
A. Z. also thanks Prof. Kleinert and the Freie Universit\"at Berlin for
their hospitality.

\newpage

\vspace*{7cm}

\noindent
{\bf Figure~1} The dynamical behaviour of the scale factors $a_{1}$ and
$a_{2}$  in harmonic time gauge for case 3. if $\nu_{1} > 0$ ($a =
2^{\frac{1}{d_{1} - 1}} a_{(0)1}$).

\vspace{9cm}
{\bf Figure~2} The dynamical behaviour of the scale factors $a_{1}$ and
$a_{2}$  in harmonic time gauge for case 3. if $\nu_{1} < 0$ ($a =
2^{\frac{1}{d_{1} - 1}} a_{(0)1}$).

\newpage


\vspace*{7cm}

\noindent
{\bf Figure~3} The dynamical behaviour of the scale factors $a_{1}$ and
$a_{2}$  in synchronous time gauge for case 3. with dimensions
$d_{1} = 3, d_{2} \ge 1$. Here we have $t_{1} = - \tilde{c}$,
$t_{2} = \tilde{c}$ and $a = 2^{1/2}a_{(0)1}$.

\end{document}